\documentclass[prd,twocolumn,superscriptaddress,showpacs,floatfix,%
preprintnumbers, nofootinbib,hyperref]{revtex4}              
\usepackage{graphicx}
\usepackage{rotating}
\usepackage{epsfig}
\usepackage[usenames,dvipsnames]{xcolor}

\usepackage{color}

\def\he4{$^4$He}
\def\h2{$^2$H}

\begin{document}

\preprint{IPPP/14/ 79, DCPT/14/ 158}

\title{Unveiling secret interactions among sterile neutrinos with big-bang nucleosynthesis}
\author{Ninetta Saviano} 
\affiliation{Institute for Particle Physics Phenomenology, Department of Physics,
Durham University,\\ Durham DH1 3LE, United Kingdom}
\author{Ofelia Pisanti}
 \affiliation{Dipartimento di Scienze Fisiche, Universit{\`a} di Napoli Federico II, Complesso Universitario di Monte S. Angelo, I-80126 Napoli, Italy}
 \affiliation{Istituto Nazionale di Fisica Nucleare - Sezione di Napoli, Complesso Universitario di Monte S. Angelo, I-80126 Napoli, Italy}
\author{Gianpiero Mangano}
\affiliation{Istituto Nazionale di Fisica Nucleare - Sezione di Napoli, Complesso Universitario di Monte S. Angelo, I-80126 Napoli, Italy}
\author{Alessandro Mirizzi} 
\affiliation{II Institut f\"ur Theoretische Physik, Universit\"at Hamburg, Luruper Chaussee 149, 22761 Hamburg, Germany}

\date{\today}

\begin{abstract}
Short-baseline neutrino anomalies suggest the existence of low-mass ($m\sim {\mathcal O}(1)$~eV) sterile neutrinos
$\nu_s$.
These would be efficiently produced in the early universe  by oscillations with active neutrino species, leading
to  a thermal population of the sterile states seemingly incompatible with cosmological observations.
In order to relieve this tension it has been recently speculated that new ``secret'' interactions among sterile neutrinos,
mediated by a massive gauge boson $X$ (with $M_X \ll M_W$), can inhibit or suppress the 
sterile neutrino thermalization, due to the production of a large matter potential term. We note however, that
they also generate strong collisional terms in the sterile neutrino sector that induce an efficient sterile neutrino production after a resonance in matter is encountered, increasing their contribution to the number of relativistic particle species $N_{\rm eff}$. Moreover, for values of the parameters of the $\nu_s$-$\nu_s$ interaction for which  the resonance  takes place
at temperature $T\lesssim$ few~MeV, significant distortions are produced in the electron (anti)neutrino spectra, altering the abundance of light element in Big Bang  Nucleosynthesis (BBN). 
Using the present determination of $^4$He and deuterium primordial abundances we determine the BBN constraints on the model parameters. We find that $^2$H/H density ratio exclude much of the parameter space if one assume a baryon density at the best fit value of Planck experiment, $\Omega_B h^2= 0.02207$, while bounds become weaker for a higher $\Omega_B h^2=0.02261$, the 95 \% C.L. upper bound of Planck. Due to the large error on its experimental determination, the helium mass fraction $Y_p$ gives no significant bounds.

\end{abstract}


\maketitle

\section{Introduction}

In recent years  
 hints for the existence of (sub)-eV sterile neutrinos $\nu_s$, mixing
with the three active species, emerged 
from
different short-baseline neutrino oscillation experiments.
Notably, 
the    ${\overline \nu}_\mu \to {\overline\nu}_e$ oscillations in
LSND~\cite{Aguilar:2001ty} and MiniBoone~\cite{AguilarArevalo:2010wv} experiments (recently constrained by the 
ICARUS experiment~\cite{Antonello:2012pq}), the ${\overline\nu}_e$ and $\nu_e$ disappearance
revealed by the Reactor Anomaly~\cite{Mention:2011rk}, and the Gallium Anomaly~\cite{Acero:2007su}  (see~\cite{Abazajian:2012ys} for a recent review) have been described in scenarios
 with one (dubbed ``3+1'') or two (``3+2'')  sterile neutrinos (see~\cite{Giunti:2013aea,Kopp:2013vaa} for the latest analysis).
 
 Low-mass sterile neutrinos would be  produced also in the early universe via oscillations of active 
neutrinos~\cite{Dolgov:2002wy}.
 The preferred range of masses and mixing angles from the laboratory anomalies would lead to a complete 
thermalization of $\nu_s$'s~\cite{Hannestad:2012ky}, in disagreement with 
recent cosmological  analysis~\cite{Archidiacono:2013xxa,Mirizzi:2013kva,Hamann:2013iba,Wyman:2013lza,Giusarma:2014zza,Archidiacono:2014apa,Leistedt:2014sia,Bergstrom:2014fqa}.
Indeed, for sterile neutrino masses of 1~eV, or larger,  tight constraints are obtained
from structure formation, affected at small-scales by
the presence of a fully thermalized massive sterile neutrino species~\cite{Hamann:2011ge}.
Furthemore,  Big Bang Nucleosynthesis (BBN)
 marginally allows for a completely thermalized sterile neutrino~\cite{Mangano:2011ar,Cooke:2013cba}.

In order to reconcile the eV sterile neutrino interpretation of the short-baseline anomalies with the cosmological
observations, the most straightforward possibility would be  to suppress  the sterile neutrino thermalization. One of the proposed mechanism~\cite{Chu:2006ua} (see also~\cite{Foot:1995bm}) is 
to consider a  primordial  asymmetry $L_\nu$ between active neutrinos and antineutrinos. 
This would  add an additional ``matter potential'' term in the active-sterile neutrino equations of motion. If sufficiently
large, this term inhibits the active-sterile flavor conversions via the in-medium suppression of the mixing angle.
In recent papers~\cite{Hannestad:2012ky,Mirizzi:2012we,Saviano:2013ktj} it has been shown that in order to achieve a sufficient suppression of the 
sterile neutrino abundance,  an asymmetry $L_\nu \gtrsim 10^{-2}$ is required. 
However, such a large value  does not seem very 
natural and moreover, it leads to non-trivial effects on BBN due to the distortions induced on active neutrino spectra, as shown in~\cite{Saviano:2013ktj}.  

More recently, in~\cite{Hannestad:2013ana,Dasgupta:2013zpn} an alternative method to suppress the sterile neutrino production has been proposed, based on the introduction of new ``secret'' self-interactions among sterile neutrinos, mediated by a massive gauge boson $X$, with $M_X \ll M_W$.   
As in the case of neutrino asymmetries, the self-interactions would generate a matter potential in the flavor
evolution equations which suppresses
the sterile neutrino abundance. If the new interaction mediator $X$ also couples to dark matter, this might also possibly relieve some of the small-scale structure problems associated with cold dark matter~\cite{Dasgupta:2013zpn,Bringmann:2013vra}
(nonstandard interactions were also introduced 
 to alleviate problems
related to cold dark matter in~\cite{Boehm:2000gq,Boehm:2003hm,Serra:2009uu,Aarssen:2012fx,Boehm:2014vja,Chu:2014lja}
and in the references therein). 

Since these new interactions involve only the sterile neutrino sector, they would evade existing limits
on secret interactions among active neutrinos~\cite{Bilenky:1999dn,Archidiacono:2013dua,Laha:2013xua}, and therefore seem apparently unconstrained.
However, as we will discuss in this paper, these interactions can be unveiled by exploiting cosmological observations. 
Indeed, when the ``secret'' matter potential becomes of the order of the active-sterile vacuum oscillation frequency, a resonance is encountered maximizing the in-medium mixing
angle~\cite{Matt}. 
This would lead to  Mikheyeev-Smirnov-Wolfenstein
 (MSW)-like  resonant flavor conversions  among active and sterile
neutrinos~\cite{Matt}. Moreover,
 the presence of strong collisional effects in the sterile neutrino sector, again due to the secret interactions,
 on one side 
would damp the MSW conversions, on the other would \emph{enhance} the sterile neutrino production
via  non-resonant processes~\cite{Kainulainen:1990ds}.
In particular, the latter would bring the active-sterile neutrino ensemble towards the flavor 
equilibrium~\cite{Stodolsky:1986dx,Enqvist:1991qj}.

In~\cite{Hannestad:2013ana} the authors have  performed different multi-momentum simulations of the active-sterile
flavor evolution in a simplified scheme involving only one active and one sterile species.   
They found that for values of the $\nu_s$-$\nu_s$ coupling $g_X \gtrsim 10^{-2}$ and
masses $M_X \gtrsim 10$~MeV,  the resonance can produce an increase in the effective neutrino
species $N_{\rm eff}$, introduced as usual as a way to parameterize the energy density $\rho_{\rm rel}$ of relativistic species at some relevant epoch, in terms of photon contribution $\rho_\gamma$
\begin{equation}
\rho_{\rm rel}=  \rho_\gamma \left(1+ \frac{7}{8}\, \left( \frac{4}{11} \right)^{4/3} N_{\rm eff} \right) \,  .
\label{def-neff}
\end{equation}
Moreover,   resonances occurring at $T\lesssim$ few~MeV
happen so late that significant distortions are produced in the electron (anti)neutrino spectra. Both these effects
have  a potential relevance for   the abundance of light elements BBN.

Motivated by these warnings, we investigate in details these effects to obtain  constraints on the  secret $\nu_s$-$\nu_s$ interactions parameter space.
As in~\cite{Hannestad:2013ana}, we work in a  situation in which active-sterile flavor conversions
would occur at a temperature $T \ll M_X$. This implies that one can reduce 
the interaction to an effective four-fermion point-like structure, with 
strength 
\footnote{The numerical factor $\sqrt{2}/8$ has been included in order to exactly mimic the relation between 
the Fermi constant $G_F$, the $SU(2)$ coupling constant $g$ and the $W$-mass  in the Standard Model.} 
\begin{equation}
G_X = \frac{\sqrt{2}}{8} \frac{g_X^2}{M_X^2} \, .
\end{equation}
Differently from~\cite{Hannestad:2013ana} we will work in a (2+1) scenario that allows to describe more realistically the
flavor dynamics. However,  in this case computing reliably
the spectral distortions and $N_{\rm eff}$
as functions
of the secret interaction parameters is a very challenging task, involving time consuming numerical calculations for the
flavor evolution. Therefore, we will apply an averaged--momentum approximation, where 
all neutrinos share the mean thermal momentum. In this limit, the information on the active neutrino distribution distortion is contained in a single (time evolving) parameter, which weights the usual Fermi-Dirac distribution. In other terms, neutrinos will be characterized by a gray-body distribution.
 
This article is structured as follows: in Section~II we present the formalism to study 
the flavor conversions of active-sterile neutrinos in the presence 
of secret $\nu_s$-$\nu_s$ self-interactions and we show some examples of the flavor evolution. We present also our results for the value of $N_{\rm eff}$ as function of $G_X$ and $g_X$, the parameters characterizing the strength of the new interaction. 
In Section~III  we discuss the impact on BBN, in particular on primordial abundances of $^4$He and $^2$H and discuss the constraints that present experimental data on these two nuclei yields put on the secret interaction scenario.
Finally, in Section~IV we conclude.

\section{Setup of the Flavor evolution}

\subsection{Equations of motion}

In this Section we summarize  the equations of motion  for the  (2+1) active-sterile neutrino system in the early universe, using 
 the same 
notation of~\cite{Mirizzi:2012we}, to which we address the reader  for details. 
 In order to take into account the interplay between oscillations and collisions of neutrinos, it is necessary to 
describe  the neutrino (antineutrino) system in terms of  $3\times 3$ density matrices $\varrho$  ($\bar\varrho$)\footnote{Here $\nu_{\mu}$ refers generically to a non-electron active  flavor state.}
\begin{equation}
\varrho_{\bf p} =
\left(\begin{array}{ccc} 
\varrho_{ee} &  \varrho_{e \mu} & \varrho_{es} \\
\varrho_{\mu e}  & \varrho_{\mu \mu} &  \varrho_{\mu s} \\
\varrho_{s e} &\varrho_{s \mu} &\varrho_{ss}
\end{array}\right) \, . 
\label{eq:rho}
\end{equation}
Since our aim is to perform an extensive scan of the parameter space of the $\nu_s$-$\nu_s$ secret interactions,
 in order to carry out a more treatable numerical analysis,  we will consider the averaged-momentum approximation, based on the ansatz
$ \varrho_{\bf p} (T) \to f_{FD}(p)\,\rho(T)\, $ (see~\cite{Mirizzi:2012we}),
where $\rho(T)$ is  the density matrix for the mean thermal momentum $\langle p \rangle = 3.15~T$, and
 $f_{FD}(p)$ is the Fermi-Dirac neutrino equilibrium distribution, and similarly for antineutrinos. 

The evolution equation  for  the momentum-averaged density matrix $\rho$ is the following~\cite{Sigl:1992fn,McKellar:1992ja,Mirizzi:2012we}
\begin{equation}
{\rm i}\,\frac{d\rho}{dt} =[{\sf\Omega},\rho]+ C[\rho]\, ,
\label{drhodt}
\end{equation}
and a similar expression holds for the antineutrino matrix $\bar\rho$.
The evolution in terms of the comoving observer proper time $t$ can be easily recast in function of the photon temperature $T$ 
(see~\cite{Mirizzi:2012we} for a detailed treatment). 
The first term on the right-hand side of Eq.\ (\ref{drhodt}) describes the flavor oscillations Hamiltonian, given by

\begin{eqnarray}
{\sf\Omega}&=&\frac{{\sf M}^2}{2} \left\langle \frac{1}{p} \right\rangle +
\sqrt{2}\,G_{\rm F}\left[-\frac{8 \langle p \rangle}{3 }\, \bigg(\frac{{\sf E_\ell}}{M_{\rm W}^2} + \frac{{\sf E_\nu}}{M_{\rm Z}^2}\bigg)
+{\sf N_\nu}\right] \nonumber \\
&+& \sqrt{2}\,G_{\rm X}\left[-\frac{8 \langle p \rangle {\sf E_s}}{3 M_{\rm X}^2}
+{\sf N_s}\right]  \,\ ,
\label{eq:omega}
\end{eqnarray}
where ${\sf M}^2$ = ${\mathcal U}^{\dagger} {\mathcal M}^2 {\mathcal U}$ is the neutrino mass matrix.
Here 
${\mathcal U}={\mathcal U}(\theta_{e\mu}, \theta_{es}, \theta_{\mu s})$
is the $3 \times 3$ active-sterile mixing matrix, parametrized as in~\cite{Mirizzi:2012we}.  We assume $\theta_{e\mu}$ equal to the active $1-3$ 
mixing angle $\theta_{13}$~\cite{Capozzi:2013csa},   while we fix the active-sterile mixing angles to the best-fit values of the different 
anomalies~\cite{Giunti:2013aea}, namely
\begin{eqnarray}
\sin^2  \theta_{e\mu} &=& 0.023  \,\ , \\
\sin^2 \theta_{es}  &=& 0.033 \,\ ,  \\
\sin^2 \theta_{\mu s}  &=& 0.012 \,\ .
\label{eq:stermix}
\end{eqnarray}
The mass--squared matrix  
\begin{equation}{\mathcal M}^2 = \textrm{diag}(-\Delta m^2_{\rm atm}/2,
+\Delta m^2_{\rm atm}/2,\Delta m^2_{\rm st}) 
\end{equation}
 is parametrized in terms of the atmospheric mass-squared difference
$\Delta m^2_{\rm atm}= 2.43 \times 10^{-3}$~eV$^2$~\cite{Capozzi:2013csa} and of the active-sterile mass splitting 
$\Delta m^2_{\rm st}=1.6$~eV$^2$, fixed from the short-baseline fit in 3+1 model~\cite{Giunti:2013aea}. In the following
we assume the normal mass hierarchy $\Delta m^2_{\rm atm} >0$.

The terms proportional to the Fermi constant $G_F$ in Eq.~(\ref{eq:omega})   encode the standard matter effects in the neutrino oscillations.
In particular, the term ${\sf E_\ell}$ is related to the energy density of $e^{\pm}$ pairs, 
${\sf E_\nu}$ to the energy density of $\nu$ and  $\bar\nu$,  and ${\sf N_\nu}$ 
is the $\nu-\nu$ interaction term  proportional to the neutrino asymmetry.
The term proportional to $G_X$ represents the new matter secret potential 
where ${\sf E_s}$ is  the energy density of $\nu_s$ and  $\bar\nu_s$
\footnote{See~\cite{Dasgupta:2013zpn} for an explicit calculation of the neutrino potential associated with
the secret interactions.}.
This refractive term can induce a MSW-like resonance between the
active and sterile states when 
it becomes of the same order of the vacuum  frequency associated with the active-sterile mass-squared splitting~\cite{Hannestad:2013ana,Dasgupta:2013zpn}.
Finally, ${\sf N_s}$ is the self-interaction term proportional to the sterile neutrino asymmetry.  
In the following, we will consider the most conservative scenario, with zero neutrino asymmetries in both the active and sterile
sectors, so that $\bar\rho=\rho$. 

The last term in the right-hand side of Eq.~(\ref{drhodt}) is the collisional term.
Following~\cite{Mirizzi:2012we}, for the Standard Model interactions we write it as
\begin{eqnarray}
{C}_{\rm SM}[{\rho}] &=& -\frac{i}{2} G_F^2 \,
 (\{{\sf S}^2, \rho-{\sf I}\} 
 - 2 {\sf S}(\rho-{\sf I}){\sf S} \nonumber \\
&+& \{{\sf A}^2,
 (\rho-{\sf I})\} + 2 {\sf A}(\bar\rho-{\sf I}){\sf A}) \,\ . \nonumber\\
 \label{eq:collision}
 \end{eqnarray}
In flavor space,  ${\sf S} = \textrm{diag}(g_s^e, g_s^\mu,0)$ and 
 ${\sf A} = \textrm{diag}(g_a^e, g_a^\mu,0)$ with    the numerical coefficients for the scattering and annihilation processes of the different flavors (see~\cite{Mirizzi:2012we} for the numerical values). 
 Concerning the collisional term associated with the secret interactions, one should change $G_F \to G_X$ 
in Eq.~(\ref{eq:collision}), and introduce two new matrices of coefficients
${\sf A}_X$ and ${\sf S}_X$ for annihilations and scatterings mediated by the gauge boson $X$, respectively. Since we consider here masses of the new boson larger than MeV, in the relevant temperature range $M_X \gg T$ we can neglect annihilation processes and assume ${\sf A}_X= 0$. For the scattering
matrix we have ${\sf S}_X = \textrm{diag}(0, 0,1)$. 
Therefore, we get the following expression
\begin{equation}
{C}_{\rm X}[{\rho}] = -\frac{i}{2} G_X^2 \,
 (\{{\sf S}_X^2, \rho-{\sf I}\} 
 - 2 {\sf S}_X(\rho-{\sf I}){\sf S}_X) \,\ . \\
 \label{eq:collisionnsi}
 \end{equation}

 The strong collisional effects produce a damping of the resonant transitions and would bring the system
towards the flavor equilibrium among the different neutrino species  with
a rate 
$\Gamma_t \sim \sin^2 2\theta_{m} G_X^2$~\cite{Stodolsky:1986dx,Kainulainen:1990ds,Enqvist:1991qj}, where $\theta_{m}$ is the in-medium mixing~\cite{Kuo:1989qe}.
At resonance $\theta_{m} \simeq \pi/4$, so that  both the resonant and non-resonant $\nu_s$ production are maximized.

To conclude, we mention that in our average-momentum treatment we cannot account for a  redistribution in the energy
spectra of $\nu_s$ associated with the elastic scatterings, considered in the multi-momentum
treatment of~\cite{Hannestad:2013ana}. However, this effect would not be expected to  produce a major impact on our results, described in Section III. Though more refined bounds on  the allowed values of $g_X$ and $G_X$ could be only obtained with a full multi--momentum analysis, the main conclusion, basically that there is a strong tension between the secret interaction scenario and primordial deuterium yield, would not change drastically. 
Finally, we also stress that a precision computation seems also
premature and illusory, given the dependence from unknown or poorly constrained parameters in the active sterile mixing from short--baseline neutrino anomalies such as, for example the $\nu_s-\nu_\tau$ mixing.

\begin{figure*}[!t]
   \includegraphics[angle=0,width=1.\columnwidth]{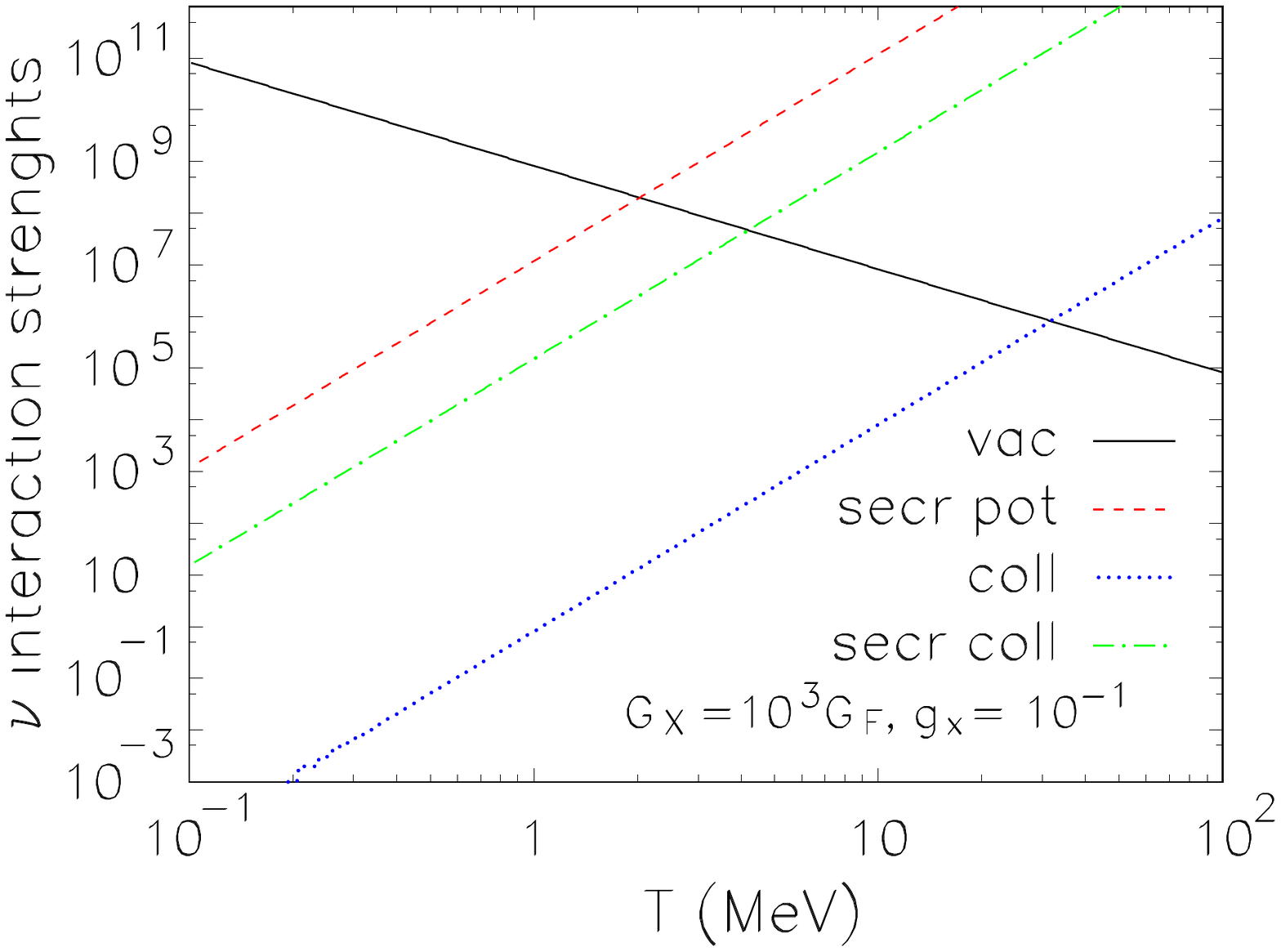} 
  \includegraphics[angle=0,width=1.\columnwidth]{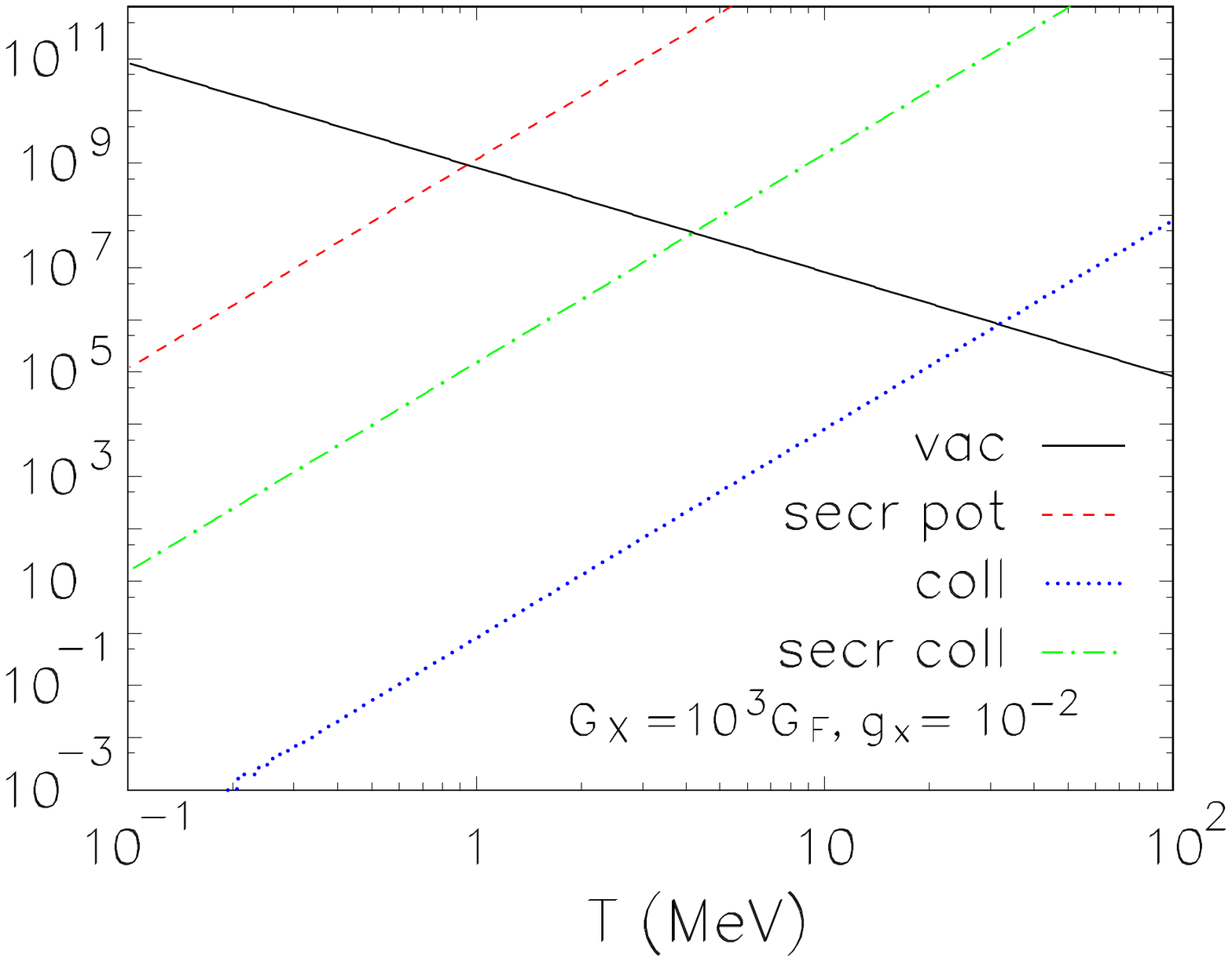} 
  \vspace{-3.5cm}
\caption{Neutrino  refractive and collisional rates (normalized in terms of the Hubble rate) versus temperature $T$ for $G_X=10^3$~$G_F$. Left panel corresponds to $g_X=10^{-1}$, while right panel to
$g_X=10^{-2}$. 
The curves correspond to the active-sterile vacuum  term (solid curve), the secret matter potential
for $\rho_{ss}=10^{-2}$  (dashed curve), 
the standard collisional term (dotted curve) and the collisional term associated with $G_X^2$ (dot-dashed curve)
\label{fig1}.} 
\end{figure*}
 
\begin{figure*}[!t]
\begin{center}
  \includegraphics[angle=0,width=0.5\textwidth]{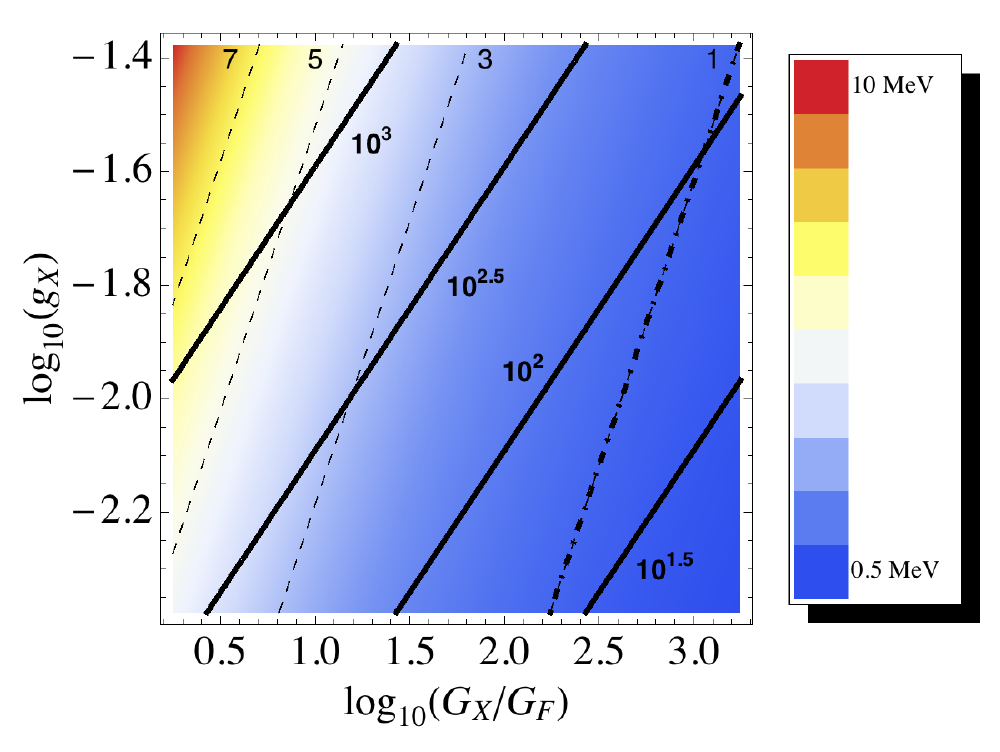} 
    \end{center}
\caption{Resonance temperature $T_{\rm res}$ in the plane  $(G_X, g_X)$. 
Dashed curves represent constant $T_{\rm res}$ contours, while on solid curves
$M_X$ is constant. The values shown for both parameters are expressed in MeV. The case of $T_{\rm res}=1 $ MeV is highlighted with a tick dot-dashed style, and correspond to the order of magnitude of BBN onset in the standard case, when neutron to proton density ratio freezes out. \label{fig2}} 
\end{figure*}

\subsection{Sterile neutrino  production}
 
In Fig.~\ref{fig1} we show the behavior of the different neutrino  refractive and collisional rates normalized to the Hubble rate, versus photon temperature $T$
(see~\cite{Mirizzi:2012we} for details).
Results are shown for $G_X=10^3$~$G_F$. In the left panel we consider $g_X=10^{-1}$
while in the right panel panel we take $g_X=10^{-2}$, corresponding to  $M_X= 390$~MeV and 
 $M_X= 39$~MeV, respectively. In  these plots we show the active-sterile vacuum  term (solid curve) and the secret matter potential 
assuming $\rho_{ss}=10^{-2}$ (dashed curve), an indicative value when steriles are about to be excited.
We realize that for $g_X =10^{-1}$ the  active-sterile resonance  occurs at $T\simeq 2$~MeV, while for
 $g_X =10^{-2}$ it takes place at  $T\simeq 1$~MeV. We also show
the standard collisional term (dotted) and the collisional term associated with $G_X^2$ (dot-dashed curve).
Notice that in the active sector the system remains collisional down to a few MeV, when the standard collision rate over Hubble rate drops below unity.
On the other hand, the secret collisional term associated with $G_X^2$  remains larger than the Hubble rate till $T\sim 0.1$~MeV, implying that
it will tend to drive the active-sterile system towards the flavor equilibrium for all the relevant evolution. 
The  secret collisional term also dominates over the vacuum oscillation one for  $T\gtrsim$ few~MeV, thus breaking the coherence between different  neutrino flavors and preventing significant oscillations at high temperatures.

\begin{figure*}[!t]
  \includegraphics[angle=0,width=0.7\textwidth]{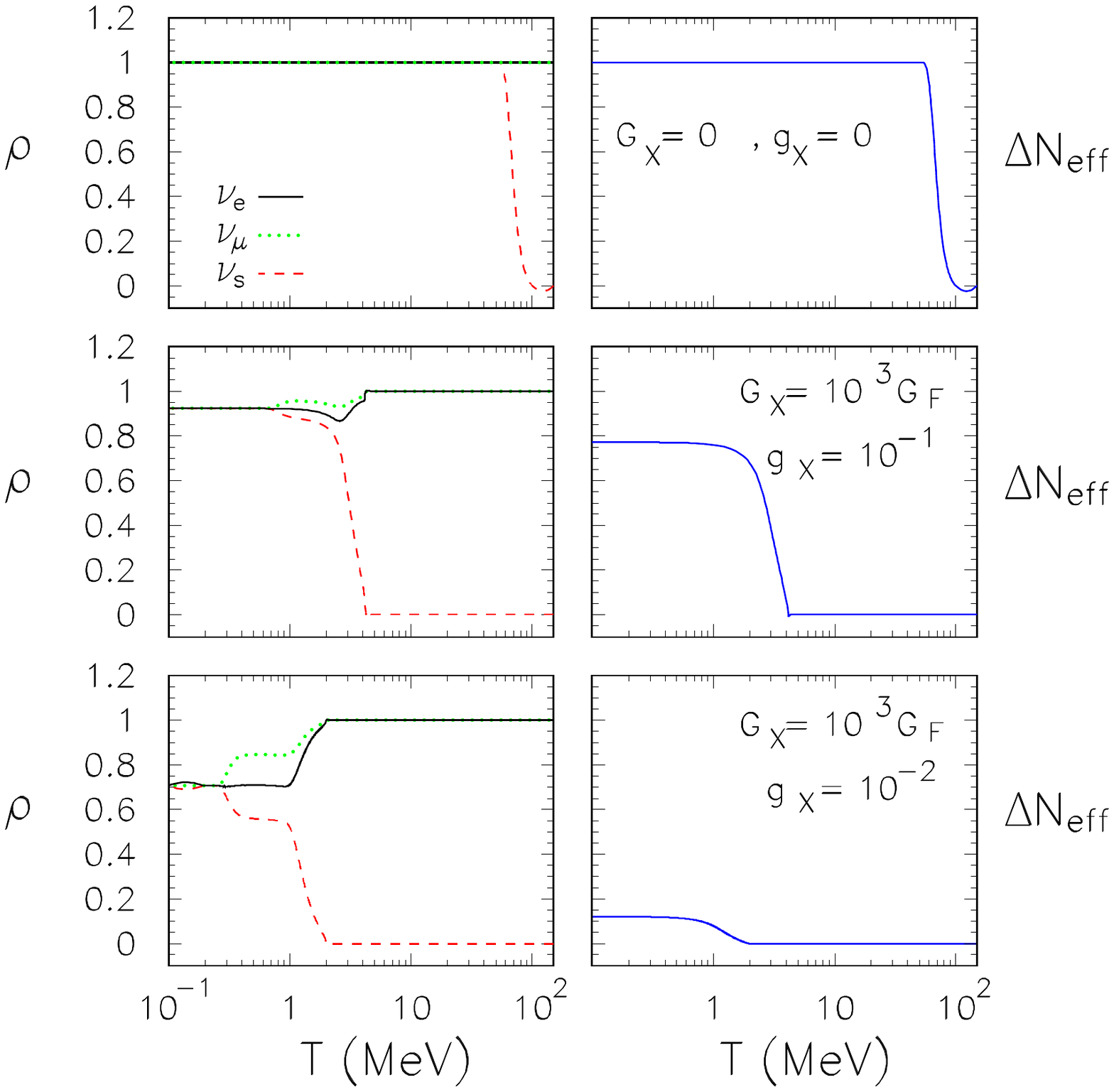} 
 \vspace{-3.5cm}
\caption{Flavor evolution as functions of temperature $T$ for different cases for  $G_X=10^{3}$~$G_F$. Upper panel is the standard case without secret interactions. Middle and lower plots are for $g_X=10^{-1}$ and $g_X=10^{-2}$, respectively. 
In the left panels we report  { $\rho_{ee}$  (continuous curve), $\rho_{\mu\mu}$ (dotted curve)
and $\rho_{ss}$ (dashed curve)}. Right panels show the corresponding $\Delta N_{\rm eff}$. 
 \label{fig3}} 
\end{figure*}

In Fig.~\ref{fig2} we show the resonance temperature $T_{\rm res}$ for $\rho_{ss}=10^{-2}$ in the plane
 $(G_X, g_X)$.
.  From the resonance condition, one gets 
\begin{equation}
\frac{T_{\rm res}}{\textrm{MeV}}  \simeq 0.4 \left[\bigg(\frac{\Delta m^2_{\rm st}}{\textrm{eV}^2} \bigg)
\bigg(\frac{M_X^2}{\textrm{MeV}^2}\bigg) \bigg(\frac{G_F}{G_X}\bigg) \frac{1}{\rho_{ss}}\right]^{1/6} \,\ . 
\end{equation}
Dashed and solid curves represent locations of constant $T_{\rm res}$ and $M_X$, respectively.  We see that for the values of the parameters of the secret interactions
considered in this figure, resonances may take place in a range of temperatures relevant for BBN. In the following we will focus on this range of parameters for our analysis.

\begin{figure*}[!t]
  \includegraphics[angle=0,width=1.\columnwidth]{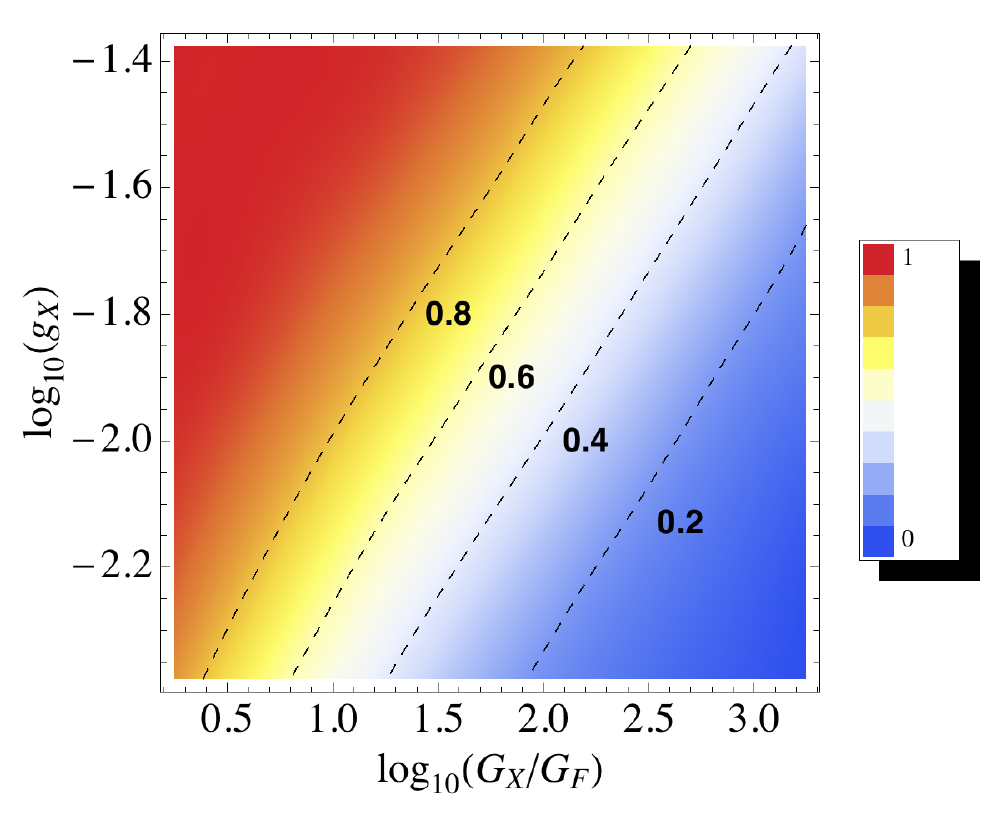} 
\caption{ The asymptotic values of $\Delta N_{\rm eff}$ versus $G_X$ and $g_X$. Colours from blue (lower right corner) to red (upper left corner) correspond to increasing values. Dashed curves show some reference values. \label{fig4}} 
\end{figure*}

We now consider the temperature evolution of neutrino momentum--averaged density matrix $\rho(T)$. In Fig.~\ref{fig3} we show the flavor evolution for the two cases of Fig.~\ref{fig1} compared to the standard case without secret interactions. In particular, 
in the left panels we represent $\rho_{ee}$  {(continuous curve), $\rho_{\mu\mu}$ (dotted curve)
and $\rho_{ss}$ (dashed curve)} while right panels show the corresponding $N_{\rm eff}$. The upper panel
is the standard evolution in absence of sterile self--interactions; in this case sterile neutrinos are produced efficiently at $T \simeq 70$~MeV and they thermalize, giving $N_{\rm eff}=4$. 
Since the sterile production occurs at temperatures for which the active neutrinos are in a collisional regime, the latter
are efficiently repopulated with $\rho_{ee}=\rho_{\mu\mu}=1$. 
In the presence of secret $\nu_s$-$\nu_s$ interactions,  
the sterile neutrino production is suppressed with respect to the standard case, till a resonance is intercepted.
For the case of $g_X=10^{-1}$ the production starts around $T =3$~MeV. Due to a strong collisional term in the sterile neutrino sector,
the system tends to evolve towards a flavor equilibrium with equal density of actives and sterile species. In this  case since the flavor conversions occur in a  temperature range where the active neutrinos are still in a collisional regime, though close to decoupling, this
effect tend to repopulate the active sector, producing a final $\rho_{ee} \simeq 0.9$ and $\Delta N_{\rm eff} =0.8$. Finally, for  $g_X=10^{-2}$ active-sterile neutrino conversions occur at $T\lesssim 1$~MeV, where active neutrinos are no longer repopulated by the collisional effects. In this case, the depletion of the $\rho_{ee}$
is remarkable, with a final value $\rho_{ee}= 0.7$ and $\Delta N_{\rm eff}=0.18$.

 We have performed a scan of neutrino evolution as function of $G_X$ and $g_X$ focusing, as already mentioned,  on the range where sterile productions is expected may alter the results of BBN. Our findings are summarized in Fig.~\ref{fig4}, where we report the asymptotic values of $\Delta N_{\rm eff}$ in the $(G_X,g_X)$ plane.  Some reference values are shown as dashed lines.

\section{Impact on observables: the light nuclei abundances}\label{BBN}

\subsection{Data and analysis}
As known, BBN proceed in two basic steps. At the MeV scale weak processes maintaning chemical equilibrium between neutrons and protons become ineffective, when their rates drop below the value of the Hubble expansion rate. Later on, at $T \sim 80$ keV deuterium forms, and soon the whole nuclear reaction chain starts till it eventually stops at a temperature of order 10 keV. The role of neutrinos, including possible sterile states, is twofold. They gravitate and contribute as relativistic species to the total energy budget. This effect is encoded into a  single parameter, $N_{\rm eff}$. In addition, electron neutrino distribution in phase space is a key input for the weak proton--neutron rates, which fix the n/p ratio. 

In the following we will exploit both the abundances of $^4$He, typically cast in terms of its mass fraction $Y_p$, and the deuterium to hydrogen number density ratio, $^2$H/H. In fact, for both these nuclei we have trustable experimental determinations of their primordial values, which we will very briefly discuss later.

We start by summarizing how typically non standard neutrino abundances influence final light nuclei  yields. For a fixed baryon density an increase of $\Delta N_{\rm eff}$ shifts the weak rate freeze out at larger temperatures, since the Hubble rate is proportionally larger. From chemical equilibrium, which we can trust down to the freezing temperature 
$T_{\rm fr}$, n/p $\sim \exp\left(-\Delta m/T_{\rm fr}\right)$, with $\Delta m$ the neutron--proton mass difference. If $T_{\rm fr}$ gets higher due to an increase of $N_{\rm eff}$, there are more neutrons available at the onset of deuterium formation, and this translates into a larger value of $Y_p$. This is what we expect if sterile neutrino states are excited in the early universe. This effect can be compensated if electron neutrino number density is {\it larger}, since in this case weak rates  get increased. The secret interaction scenario, while may reduce the value of $\Delta N_{\rm eff}$ to some (positive) values smaller than unity yet, it typically leads to a {\it smaller} electron neutrino density $\rho_{ee}<1$
These two features, i.e. a positive $\Delta N_{\rm eff}$ and less electron neutrinos, both conspire to produce a larger $Y_p$. 

The way deuterium changes is less obvious. For a larger $N_{\rm eff}$ but keeping constant the $\nu_e$ distribution, the ratio $^2$H/H increases. Nevertheless, in the case at hand it is quite difficult to get a simple feeling of how it would change depending on the two parameters $G_X$ and $g_X$.

To exploit in a quantitative manner the BBN predictions for light nuclei, we have implemented the secret interaction scenario in the public numerical code  \texttt{PArthENoPE} \cite{Pisanti:2007hk}. The value of $\Delta N_{\rm eff}$ obtained by solving the equations of motion in Eq. (\ref{drhodt}) as function of the photon temperature is now read  as an external input by \texttt{PArthENoPE}, as well as the electron neutrino distribution which is used to compute the weak thermal rates of the neutron/proton reactions. We recall that neutrino distributions are evolved in the average--momentum approximation, so the $\nu_e$ distribution used in the thermal rate is the standard Fermi-Dirac function times $\rho_{ee}(T)$. Actually, at the level of approximation we are interested in, we compute the tree level Born weak rates with the modified electron neutrino distribution instead of using the standard rates employed in the public version of the code, which also contains the contribution of radiative corrections. The latter are quite involved function depending also on neutrino distribution, which should be in principle recomputed. The approximation we are using is thus to assume that the effect of a different $\nu_e$ distribution in the one-loop contribution  rescales in the same way the Born rates do. We recall that radiative corrections contribute typically for 4-5 \% of the Born rates around the freeze out temperature, see e.g. \cite{Esposito:1998rc}. Therefore, we neglect them.

In the analysis we have used the last updated result on $Y_p$ reported in \cite{Aver:2013wba}, based on a regression
to zero metallicity with new He I emissivities, and using the dataset of \cite{Izotov:2007ed}
\begin{equation}
Y_p = 0.2465 \pm 0.0097 \, ,
\label{he4exp}
\end{equation}
while for deuterium we consider the recent result of \cite{Cooke:2013cba}
\begin{equation}
^2\mbox{H}/\mbox{H} = ( 2.53 \pm 0.04 ) \times 10^{-5} \, .
\label{h2exp}
\end{equation}
This value is the result of a reanalysis of all known deuterium absorption systems, including the new discovered very metal--poor damped Lyman--$\alpha$ system at redshift  $z=3.06726$ toward the QSO SDSS J1358+6522. Notice the quite small uncertainty, of the order of 1.6 \%.
 
Before discussing our findings we make a last remark. While the uncertainty on the theoretical value of $^4$He from \texttt{PArthENoPE} is extremely small\footnote{The only source of uncertainty is in fact, the small error on neutron lifetime, $\tau_n = 880.0 \pm 0.9$ sec \cite{Beringer:1900zz}.} and negligible with respect ot the experimental uncertainty of Eq. (\ref{he4exp}), the prediction for deuterium abundance is still affected by a large error (mainly) due to the present uncertainty on the astrophysical factor of the $d(p,\gamma)^3$He reaction, which is the leading destruction channel of this nuclide. Once we propagate the uncertainty reported in~\cite{Adelberger:2010qa}, see also \cite{Serpico:2004gx}, the value of $^2$H/H change by the amount $\pm 0.06 \times 10^{-5}$, which is even larger than the experimental error of Eq.~(\ref{h2exp}). A theoretical {\it ab initio} calculation of the S-factor for this process is also available, which suggests a higher cross section in the center of mass energy range relevant for BBN  \cite{Viviani:1999us,Marcucci:2004sq,Marcucci:2005zc}, whose impact has been recently analyzed in 
\cite{Nollett:2011aa,DiValentino:2014cta}. The predicted value of deuterium is in this case lower than if we used the experimental best fit of the rate, and in a better agreement with the experimental result of \cite{Cooke:2013cba}. In order to have a clear assessment of the error budget on deuterium theoretical prediction, we have decided to conservatively use the present experimental results of \cite{Adelberger:2010qa}, but we stress that it would be important also for the issue considered in this paper to have new experimental data on $d(p,\gamma)^3$He in the BBN energy range, with a higher precision. If the theoretical {\it ab initio} calculation would be confirmed, and the experimental error on the astrophysical factor of the $d(p,\gamma)^3$He reaction would be reduced by say, a factor three, which seems plausible \cite{DiValentino:2014cta}, the constraints we will discuss later would become more stringent.

\begin{figure*}[!t]
  \includegraphics[angle=0,width=1.\columnwidth]{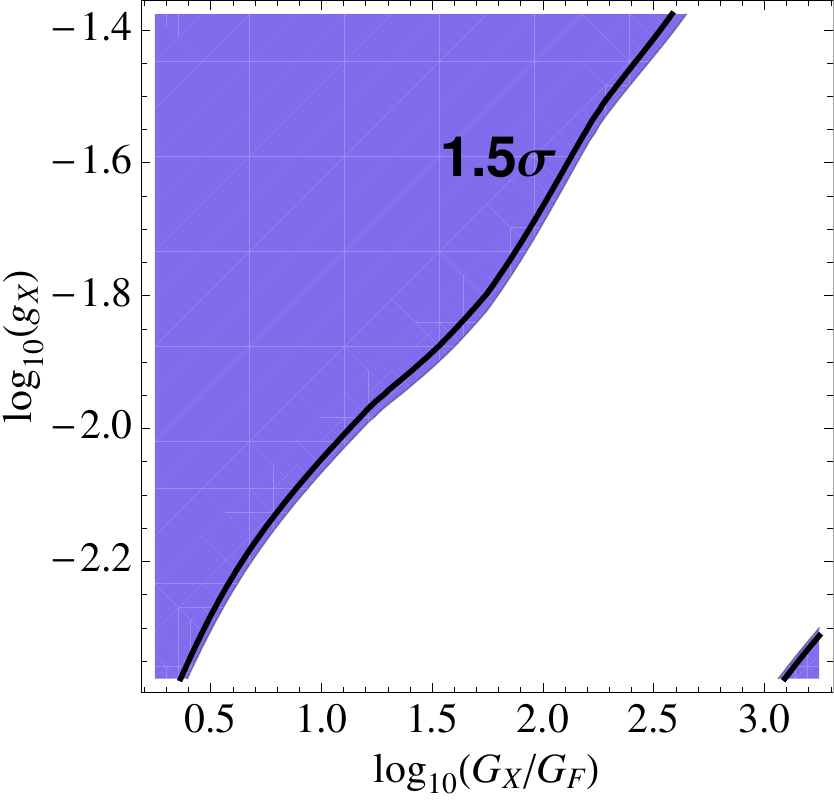} 
\caption{${}^{4}$He results. The dark region is the $1.5\sigma$ allowed parameter space for the helium mass fraction $Y_p$ using the experimental result of Eq. (\ref{he4exp}), varying the baryon density parameter in the range $0.02153 \leq \Omega_B h^2 \leq 0.02261$, corresponding to the 95 \% C.L. Planck range. The solid line bounds the permitted region if we fix  $\Omega_B h^2 = 0.02207$, the best fit quoted by Planck collaboration. At $2\sigma$ the whole region shown for $G_X$ and $g_X$ {\it would be allowed}, while at $1\sigma$ it is all excluded. \label{fig5}}
\end{figure*}

\subsection{Results}
\begin{figure*}[!t]
  \includegraphics[angle=0,width=1.\columnwidth]{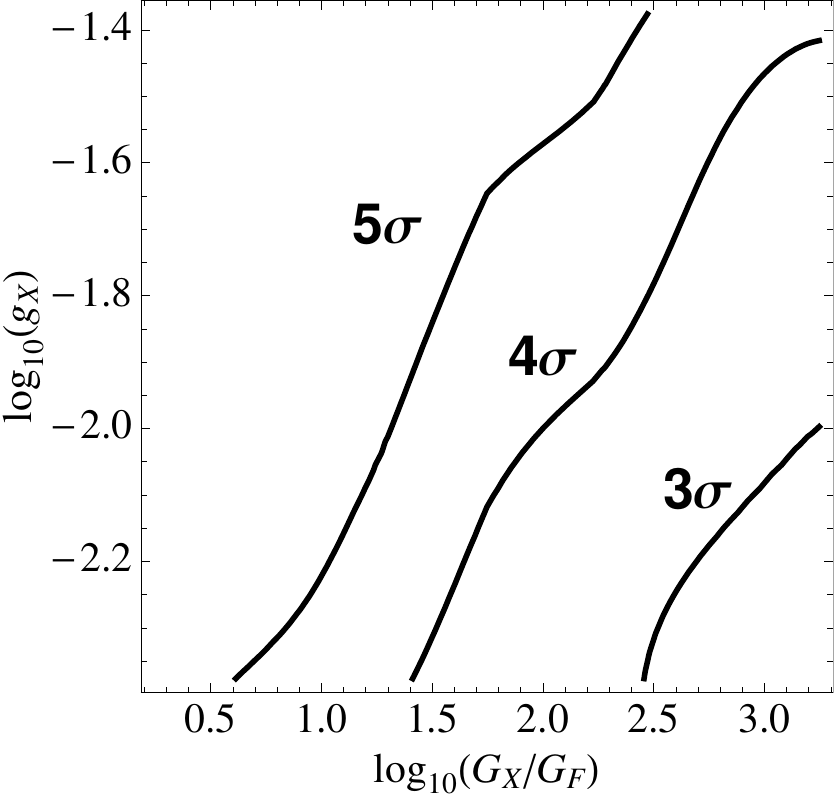} 
   \includegraphics[angle=0,width=1.\columnwidth]{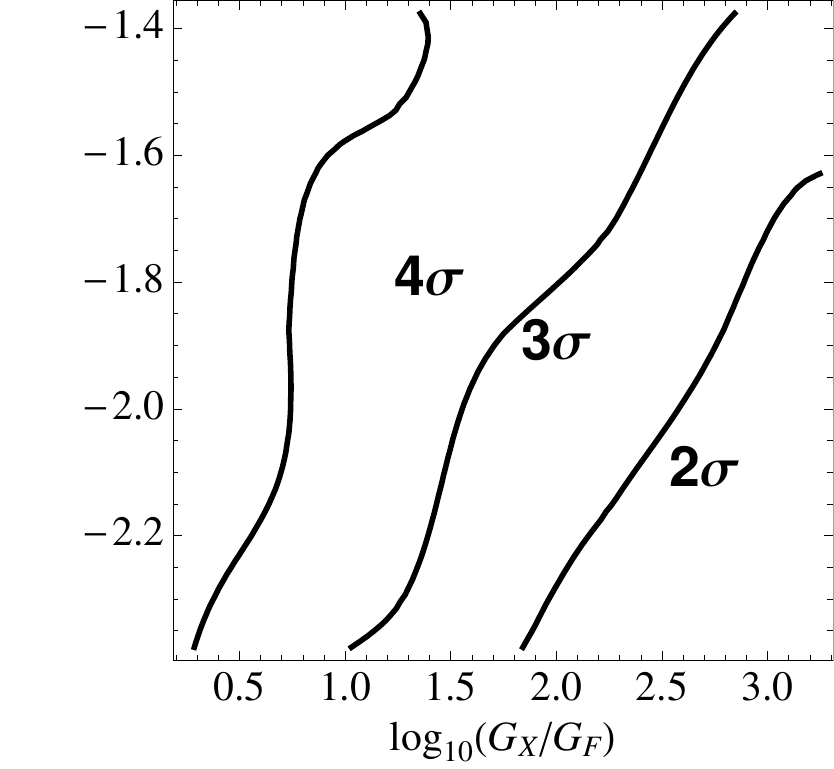} 
\caption{$^2$H/H results. Left panel: the region below each curve is the allowed one at the number of $\sigma$ shown for each case, using the experimental determination for $^2$H/H of Eq. (\ref{h2exp}) and for $\Omega_B h^2 = 0.02207$, the best fit quoted by Planck collaboration. Right panel: the same as in the left panel but for a higher baryon density $\Omega_B h^2 = 0.02261$, i.e. the Planck upper limit at 95 \% C.L.
 \label{fig6}} 
\end{figure*}

Our results are summarized in Figures~\ref{fig5} and~\ref{fig6}, showing the bounds on the secret interaction parameters 
$(G_X,g_X)$
coming from $Y_p$ and deuterium, respectively. 
By looking at  these panels 
one can  grasp the typical dependence of $^2$H/H and $Y_p$ upon the two relevant parameters. Namely, 
helium is an {\it increasing} function of $G_X$ for fixed $g_X$, while it decreases with $g_X$ for a given $G_X$.
In the same range deuterium shows exactly the opposite behavior.
Consider first the helium mass fraction $Y_p$. The two dark regions in Fig.~\ref{fig5} in the upper left and lower right corners are the allowed $G_X$ and $g_X$ at 1.5$\sigma$, where $\sigma$ is the experimental error on $Y_p$ of Eq.~(\ref{he4exp}), the theoretical error being completely negligible. The value of the baryon density is varied in the 95 \% C.L. region of Planck results, $\Omega_B h^2 = 0.02207 \pm 0.00054$~\cite{PlanckXVI}. The solid line bounds instead the 1.5$\sigma$ allowed region if we fix $\Omega_B h^2$ at the Planck best fit value. We see that for both these cases almost all the parameter space is excluded. We have shown the 1.5$\sigma$ exclusion contours,  since for the 2$\sigma$ cases the whole range for $G_X$ and $g_X$ of Fig.~\ref{fig5} {\it would have been permitted}. Conversely,  at 1$\sigma$ the whole plane would be instead {\it excluded}. In fact,  the  uncertainty on $Y_p$, at the level of 4 \%, is too large to severely constrain the secret interaction parameter, which are indeed all permitted at 2$\sigma$. Instead, as we will show later,
the deuterium constraints will be much stronger.

In order to understand the dependence of $Y_p$ on the secret interaction parameters, note that
  if we decrease $g_X$ for a given $G_X$, the electron neutrino distribution becomes smaller, when the production takes place at smaller temperatures, where electron neutrinos are less efficiently repopulated by pair creation processes
(see e.g. Fig.~\ref{fig2}). It follows that also weak rates decrease. At the same time, decreasing $g_X$ also $\Delta N_{\rm eff}$ becomes smaller and thus the expansion rate is also lowered. The combination of two effects  causes an increase in $Y_p$  when lowering $g_X$. To get the same value of $Y_p$ it is thus, necessary to decrease the value of $G_X$, and this explain the bending of the iso-helium contours, as the one depicted in the upper left corner of Fig.~\ref{fig5}.

These considerations only apply as long as the resonance temperature is larger or of the order of the typical scale of n/p ratio freezing. If sterile neutrinos are produced below this scale and electron neutrino distribution is still almost unchanged at say, $T\simeq 0.8$~MeV, BBN start being insensitive to the presence of sterile secret interactions. This explains why we can see another branch of allowed values for $G_X$ and $g_X$ in the lower right corner of Fig.~\ref{fig5}: in this region resonance temperature is of order 0.6--0.7 MeV, and $Y_p$ is less affected.
 
We discuss now the bound from  $^2$H/H. In Fig.~\ref{fig6} the regions below the various curves are the allowed ranges for the two parameters at a number of $\sigma$ reported for each curve. With 
$\sigma$ here we denote the total error obtained by summing in quadrature the experimental error of Eq. (\ref{h2exp}) and the one due to nuclear rate uncertainties~\cite{Adelberger:2010qa}. From the left panel of the Figure we see that if baryon density is fixed at the Planck best fit, almost all the range shown for $G_X$ and $g_X$ is excluded at 3$\sigma$, but the high--$G_X$ and small--$g_X$ lower right corner. 
If we translate the bound in terms of the $X$ boson mass, we obtain at 3$\sigma$ as allowed mass range 
\begin{equation}
 M_X \leq 40~\textrm{MeV} \,\ . 
\end{equation}

Notice  that in absence of sterile neutrinos the  \texttt{PArthENoPE} code prediction for $^2$H/H at $\Omega_B h^2 = 0.02207$ is $^2$H/H=$(2.65 \pm 0.07) \times 10^{-5}$, which is already larger than the experimental value we are using, and compatible with it at the 2$\sigma$ level. When switching on the $\nu_s$ secret interactions, deuterium relic abundance is strongly enhanced, unless we consider the low--$g_X$ and high--$G_X$ range, the lower-right corner of Fig.~\ref{fig6}, where now the agreement with experimental value can be at least at the level of 3$\sigma$. The improvement
in this region  of the agreement
between the theoretical values of deuterium and the experimental determination of Eq.~(\ref{h2exp})
 is due to the fact that in this range of parameters the resonance temperature is below or of the order of 0.8 MeV (see Fig.~\ref{fig2}), the typical scale of n/p ratio freezing. As already noticed, in this limit BBN starts being blind to electron neutrino distortion and the only effect is the faster expansion rate due to a positive $\Delta N_{\rm eff}$, which gives a slightly more deuterium abundance. 

The constraint from $^2$H/H is weaker if we take a larger value of $\Omega_B h^2= 0.02261$, the 95 \% C.L. upper limit from Planck and the bound on the $X$ boson mass becomes $M_X \leq 220$ MeV at 3$\sigma$.
 This is because deuterium is a rapidly decreasing function of baryon fraction, so one way to compensate for a higher theoretical prediction is to shift $\Omega_B h^2$ towards larger values. In this case too, however, most of the parameter range is excluded at 2$\sigma$. We conclude that the secret interaction scenario is in tension with present data on primordial deuterium, unless baryon density is quite at the upper boundary of Planck result.

\section{Conclusions}\label{conclu}

Secret interactions among sterile neutrinos, mediated by a  gauge boson with 
$M_X \ll M_W$,  have been recently proposed~\cite{Hannestad:2013ana,Dasgupta:2013zpn}  as a possible mechanism to suppress the thermalization of eV sterile
neutrinos in the early universe, by the large matter potential they generate in this case.
However, the active-sterile neutrino ensemble would also experience a resonance 
when the matter  term gets close to the vacuum oscillation frequency. In this situation 
the sterile neutrino production would be enhanced by the combination of a resonant production, with
a non-resonant one due to the large collisional effects, caused by the secret interactions
in the sterile $\nu$ sector. 

For values of the coupling constant $g_X \gtrsim 10^{-2}$ and masses of the gauge boson 
$M_X \gtrsim {\mathcal O} (10)$~MeV~\cite{Hannestad:2013ana}, the sterile neutrino production would occur at epochs relevant
for the BBN. In this paper we have analyzed the BBN bounds on the  secret interaction scenario. The standard BBN dynamics is indeed, possibly changed by a larger value of the number of effective relativistic degrees of freedom, $N_{\rm eff}$, and the spectral distortions on $\nu_e$ induced by the active-sterile
flavor conversions. Using the present determination of $^4$He and deuterium primordial abundances, we found that due to the 4 \% error on experimental determination of helium mass fraction, $Y_p$ gives no significant bounds. 
We comment that very recently a new measurement  of $Y_p=0.2551 \pm 0.0022$ has been presented~\cite{Izotov:2014fga}. The smaller value of the quoted error would allow one to put stronger limits
than the ones we obtained.

The $^2$H/H density ratio excludes much of the parameter space if one assume a baryon density at the best fit value of Planck experiment, $\Omega_B h^2= 0.02207$. In this case we can set an upper limit on the $X$ boson mass at 3$\sigma$
$ M_X \leq 40  \,\, \textrm{MeV}\, .$
This bound   becomes weaker for a higher baryon fraction, $\Omega_B h^2=0.02261$, which is the 95 \% C.L. upper bound of Planck, that is $M_X \leq 220$ MeV. 
The bound on $M_X$ can be improved measuring the astrophysical factor of the $d(p,\gamma)^3$He process cross section in the relevant energy range, with a smaller uncertainty. 
Therefore,  new experiments measuring this quantity are therefore very welcome.


As a consequence  of our analysis the parameter space for secret interactions to reconcile sterile neutrinos with cosmology is significantly
reduced.   
 A possible way to avoid the BBN bounds is to choose the mass of the  mediator so light as $M_X \lesssim 1$~MeV
to shift the resonances at temperatures too low ($T \lesssim 0.1$~MeV) to be relevant for nucleosynthesis, as considered in~\cite{Kopp:2013vaa}. However, 
in this situation the sterile neutrino production may 
 have an effect on the cosmological neutrino 
 mass bound~\cite{Shimon:2010ug}.  Results  of this study will be presented elsewhere~\cite{Mirizzi:2014ama}.

\section*{Acknowledgements} 

We thank Silvia Pascoli for interesting discussions. N.S. also acknowledges Steen Hannestad and Thomas Tram for valuable clarifications and suggestions.
G. M., and O.P.  acknowledge support by
the {\it Istituto Nazionale di Fisica Nucleare} I.S. FA51 and the PRIN
2010 ``Fisica Astroparticellare: Neutrini ed Universo Primordiale'' of the
Italian {\it Ministero dell'Istruzione, Universit\`a e Ricerca}. The work of A.M.   was supported by the German Science Foundation (DFG)
within the Collaborative Research Center 676 ``Particles, Strings and the
Early Universe.'' 
N.S. acknowledges support from
the European Union FP7 ITN INVISIBLES (Marie Curie
Actions, PITN- GA-2011- 289442).


\end{document}